\documentclass[twocolumn,amsmath,amssymb,prl]{revtex4}
\usepackage{latexsym}
\usepackage{graphicx}
\usepackage{psfig}
\usepackage{color}

\begin{document}
\title{An apparent metal insulator transition in high mobility 2D InAs heterostructures}
\author{J.~Shabani$^{1}$}
\author{S.~Das Sarma$^{2}$}
\author{C.~J.~Palmstr{\o}m$^{1,3,4}$}
\affiliation{$^{1}$California NanoSystems Institute, University of California, Santa Barbara, CA 93106, USA
\\
$^{2}$Condensed Matter Theory Center and Joint Quantum Institute,
Department of Physics, University of Maryland, College Park, MD 2742, USA
\\
$^{3}$Department of Electrical Engineering, University of California, Santa Barbara, CA 93106, USA
\\
$^{4}$Materials Department, University of California, Santa Barbara, CA 93106, USA
}
\date{\today}
\begin{abstract}
We report on the first experimental observation of an apparent metal insulator transition in a 2D electron gas confined in an InAs quantum well. At high densities we find that the carrier mobility is limited by background charged impurities and the temperature dependence of the resistivity shows a metallic behavior with resistivity increasing with increasing temperature. At low densities we find an insulating behavior below a critical density of $n_{c} = 5 \times 10^{10}$cm$^{-2}$ with the resistivity decreasing with increasing temperature.  We analyze this transition using a percolation model arising from the failure of screening in random background charged impurities. We also examine the percolation transition experimentally by introducing remote ionized impurities at the surface. Using a bias during cool-down, we modify the screening charge at the surface which strongly affects the critical density. Our study shows that transition from a metallic to an insulating phase in our system is due to percolation transition.
\end{abstract}
\pacs{}
\maketitle
The metallic behavior of the resistivity observed at low temperatures in low-disorder two-dimensional (2D) systems is a topic of great interest in condensed matter physics. The scaling theory of localization predicts a noninteracting two-dimensional system in the presence of finite disorder is an insulator at zero temperature in the thermodynamic limit \cite{AbrahamsPRL79}. Indeed early experiments confirmed that in highly-disordered two dimensional electron systems (2DESs) the resistivity shows an insulating logarithmic temperature dependence \cite{BishopPRL80}. The scaling theory was challenged by the observation of an apparent metal-insulator transition in high-mobility electron inversion layers in Si metal-oxide-semiconductor-field-effect transistors (MOSFETs) and later in several other 2D semiconductor systems \cite{KravchenkoPRB94, AbrahamsRevModPhys01}. At higher densities, a metallic temperature dependence of the resistivity, $\rho$, and a concomitant transition to an insulating phase at lower densities were observed. This apparent metal insulator transition (MIT) is marked by a "critical carrier density", $n_{c}$ which characterizes the crossover from the higher-density metallic temperature dependence of the resistivity to the lower-density insulating temperature dependence. For $n>n_{c}$, the system exhibits a metallic behavior (d$\rho$/dT$>$0) while for $n < n_{c}$, the resistivity increases with decreasing temperature and d$\rho$/dT$<$0 in the insulating phase. 

In the past twenty years, MIT has been observed in a wide variety of 2D carrier systems such as n-Si MOSFETs  \cite{KravchenkoPRB94}, n-GaAs \cite{HaneinPRB98, LillyPRL03}, p-GaAs \cite{HaneinPRL98, ManfraPRL07, SimmonsPRL98}, n-Si/SiGe \cite{LaiAPL04, OkamotoPRB04}, p-Si/SiGe \cite{LamPRB97, ColeridgePRB97} and n-AlAs \cite{PapadakisPRB98, GunawanNatPhys07}. In the current work we present the first experimental observation of 2D MIT in a narrow gap semiconductor, namely, 2D electrons confined in InAs quantum wells.  We believe that our work is also the observation of the 2D MIT phenomenon in a material with the lowest value of the dimensionless electron interaction coupling parameter r$_{s}$ ($\sim$ 2). 



Narrow band-gap materials such as InAs are particularly interesting as they have strong spin-orbit coupling, small electron effective mass, and large g-factor. Using an electron effective band mass of $m^{*}_{e}=$ 0.023 $m_{e}$ and InAs dielectric constant $\epsilon$ = 15, the Bohr radius for InAs is $a_{B} \sim 350$ $\AA$. This is much larger than the n-GaAs system with $a_{B} \sim 100$ $\AA$ which is the next largest Bohr radius material where the 2D MIT has been reported.  By contrast, the Bohr radius in n-Si MOSFETs is around $30$ $\AA$ (and is even smaller in p-GaAs).  Since the Bohr radius is inversely proportional to the dimensionless electron interaction parameter $r_{s}$, the interaction strength in 2D InAs is typically 10 - 15 times smaller than in n-Si MOSFETs and p-GaAs for equivalent 2D carrier densities, indicating that the $r_{s}$-parameter is most likely not a decisive parameter for the existence (or not) of the 2D MIT.  The very small effective mass of InAs and its rather large Bohr radius lead to weak screening in the system compared with all other 2D systems where the 2D MIT phenomenon has so far been observed.  One way to characterize \cite{DasSarmaPRB04, DasSarmaSSC05} the strength of screening is to consider the dimensionless parameter $q_{0} =q_{TF}/2k_{F}$, where $q_{TF}$ and $k_{F}$ are respectively  the Thomas-Fermi screening wavenumber (with $q_{TF} \sim 1/a_{B} \sim m^{*}_{e}$) and the Fermi wavenumber ($k_{F} \sim n^{1/2}$ where $n$ is the 2D carrier density). It is useful to mention that $q_{0}=r_{s} g_{v}^{3/2}/\sqrt{2} \sim r_{s}$, where $r_{s}$ is the Coulomb coupling constant (i.e. the ratio of the average inter-electron Coulomb energy and the Fermi energy) and $g_{v}$ is the valley degeneracy of the 2D semiconductor ($g_{v}$=1 for our InAs system). For $q_{0} \gg 1$, we expect strong screening and consequently strong metallicity (if the system has low enough disorder) whereas for $q_{0} \ll 1$, screening is weak leading to very weak metallicity \cite{DasSarmaSSC05, DasSarmaPRB04}.  The condition $q_{0}=1$ defines a characteristic density for each system, $n_{0}$, which provides a measure of the metallicity strength (higher the value of $n_{0}$ stronger should be the metallicity).  We note that $n_{0}=1.3 \times 10^{10}$ cm$^{-2}$ (InAs), $1.6 \times 10^{11}$ cm$^{-2}$ (n-GaAs), $14 \times 10^{11}$ cm$^{-2}$ (n-Si MOSFET), $30    \times 10^{11}$ cm$^{-2}$ (p-GaAs) with the metallicity manifesting itself strongly only for $n < n_{0}$ as long as the system does not localize, i.e., $n_{0} > n_{c}$, which can only happen for samples with low disorder (i.e. high mobility).  Clearly, the metallicity is expected to be very weak in InAs (roughly 100 times weaker than in Si MOSFETs), manifesting itself at rather low densities only in rather high mobility samples. This is what we find in our InAs 2D samples:  a very weak increase of resistivity with increasing temperature which changes to a decreasing resistivity with increasing temperature at lower densities. An additional reason for the metallic temperature dependence in our 2D InAs system being extremely weak (i.e. other than the dimensionless screening parameter $q_{0}$ or the characteristic density $n_{0}$ being very small) is that the temperature range being explored is very low in the dimensionless unit $T_{0}=T/T_{F}$ \cite{DasSarmaPRB04} since the Fermi temperature $T_{F}$ for our system is $T_{F} \sim 100 \times n$ where $n$ is being measured in units of $10^{11}$ cm$^{-2}$ - the smallness of $r_{s}$, $q_{0}$, $n_{0}$, and $T_{0}$ (and consequently, the weakness of the 2D metallicity) all arise from the smallness of the InAs electron effective mass. 



A key to study the MIT in 2D systems is the ability to tune the carrier density using a top or bottom gate electrode. We note that unlike widely used GaAs systems, reliable gating has proven difficult in InAs systems due to gate leakage and hysteretic behavior. In addition, charge traps and surface Fermi level pinning could screen the applied electric field and significantly reduce the gate efficiencies. These difficulties are surmounted in the present work, in which we grow high-mobility InAs heterostructures using epitaxial growth of band engineered structures and observe, for the first time, MIT in these heterostructures.

The 2DES is realized in an unintentionally doped structure, grown on a semi-insulating InP (100) substrate using a modified VG-V80H molecular beam epitaxy system. The InAs lattice parameter has a $3.3 \%$ lattice mismatch to the InP substrate. A low temperature buffer layer is necessary to reduce and minimize the resulting dislocations in the active region. We utilize step graded In$_{x}$Al$_{1-x}$As buffer with $x=$0.52 up to 0.8 and then down to $x$=0.75 \cite{Richter00, Wallart05}. The electrons are confined to a 4 nm strained InAs layer inside an In$_{0.75}$Ga$_{0.25}$As quantum well. Relatively low electron densities could be achieved using this approach due to the different line-up of the impurity levels in InAlAs compared to InAs quantum wells grown on AlSb barriers \cite{Capotondi04, Kroemer04}. For a 110 nm-thick In$_{0.75}$Al$_{0.25}$As top barrier, a 2DES with a density of about $4 \times 10^{11}$~cm$^{-2}$ with an electron mobility of $2 \times 10^{5}$ cm$^{2}/Vs$ (mean free path $\sim 2$ $\mu$m) is measured at T = 4.2 K. The heterostructure is designed such that the first subband lies in InAs and at these densities only the first subband is occupied. The quantum well material stack and the corresponding band diagram are shown in Fig.~1. The structure is capped with 100nm-thick SiO$_{x}$.

Standard optical photolithography and wet etching techniques are used to define a mesa in a Hall bar geometry. A 100 nm-thick Ti/Au top metallic gate is deposited to cover the mesa and sidewalls. In our gated-Hall bar, we achieve full depletion at -3 V and measure less than 10 pA of leakage current. Our temperature dependence measurements were performed in a pumped $^{3}$He refrigerator where the temperature could be controlled from 0.5 K to 5 K. A standard low-frequency lock-in technique with an excitation current of 10 nA to 100 nA was used to measure both longitudinal and transverse resistances as a function of a perpendicular magnetic field. These magneto-resistances were used to determine density at each gate voltage. Examples of longitudinal magneto-resistance, $\rho_{xx}$ for several electron densities are shown in Fig.~2(b). The onset of Shubnikov-de Haas oscillations starts $\sim$ 0.5 T and Landau level filling factors vary by two for magnetic fields less than 1.5 T revealing that spin is not resolved. The measured mobilities for the corresponding traces shown in Fig.~2(b) are plotted in Fig.~2(a). This dependence is well-fitted by $\mu \sim n^{\alpha}$, with $\alpha = 0.8$ in this density range. A value of $\alpha \sim 0.8$ signifies that the mobility is limited by scattering from nearby background charged impurities~\cite{DasSarmaPRB13}. It is worth noting that mobility increases monotonically with increasing density up to our maximum experimental density $n=5 \times 10^{11}$ cm$^{-2}$.

\begin{figure}[htp]
\centering
\includegraphics[scale=0.5]{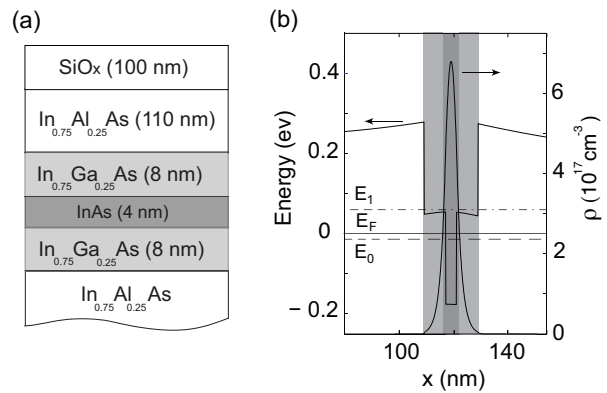}
\caption{(Color online) (a) schematic of the quantum well containing InAs layer. (b) Band diagram and charge distribution of the corresponding structure shown in (a).}
\end{figure}

\begin{figure}[htp]
\centering
\includegraphics[scale=0.9]{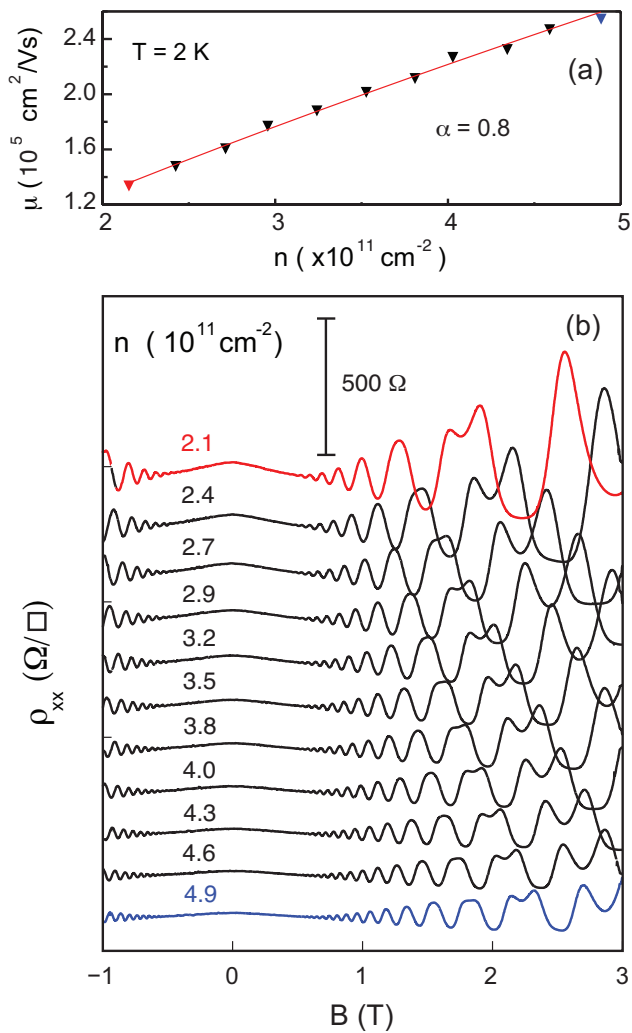}
\caption{(Color online) (a) Electron mobility plotted as a function of density (by varying the top gate bias). An exponent of $\alpha$ = 0.8 is derived from fitting the data to $\mu = n^{\alpha}$ at 2 K. (b) Corresponding magneto-resistance data measured at different densities from top to bottom: n = 2.1 (red), 2.4, 2.7, 2.9, 3.2, 3.5, 3.8, 4.0, 4.3, 4.6 and 4.9 (blue) $\times 10^{11}$ cm$^{-2}$. The lowest (and highest) density are marked in red (and blue) in (a) and (b).}
\end{figure}

\begin{figure}[htp]
\centering
\includegraphics[scale=0.6]{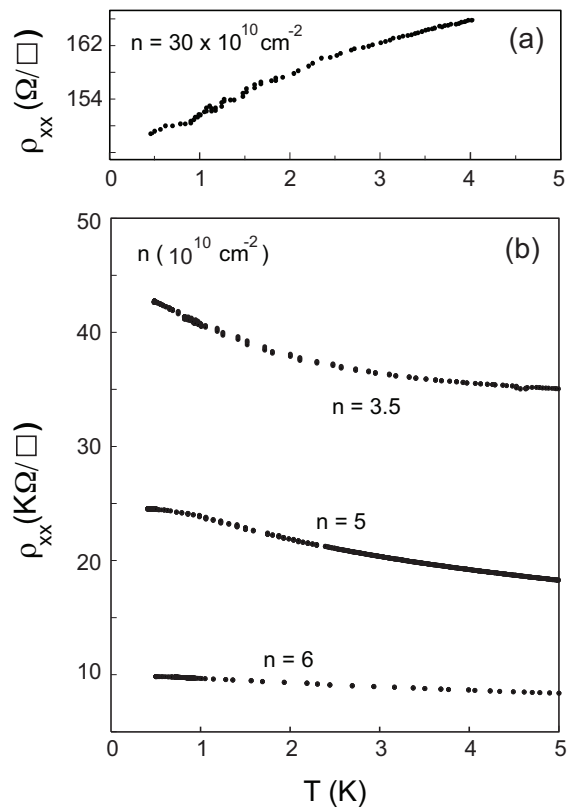}
\caption{Resistivity vs temperature at (a) $n = 30 \times 10^{10}$ cm$^{-2}$. A metallic behavior is observed. (b) For n = 3.5, 5 and 6 $\times 10^{10}$ cm$^{-2}$resistivity increases with decreasing temperature.}
\end{figure}
\begin{figure}[htp]
\centering
\includegraphics[scale=0.55]{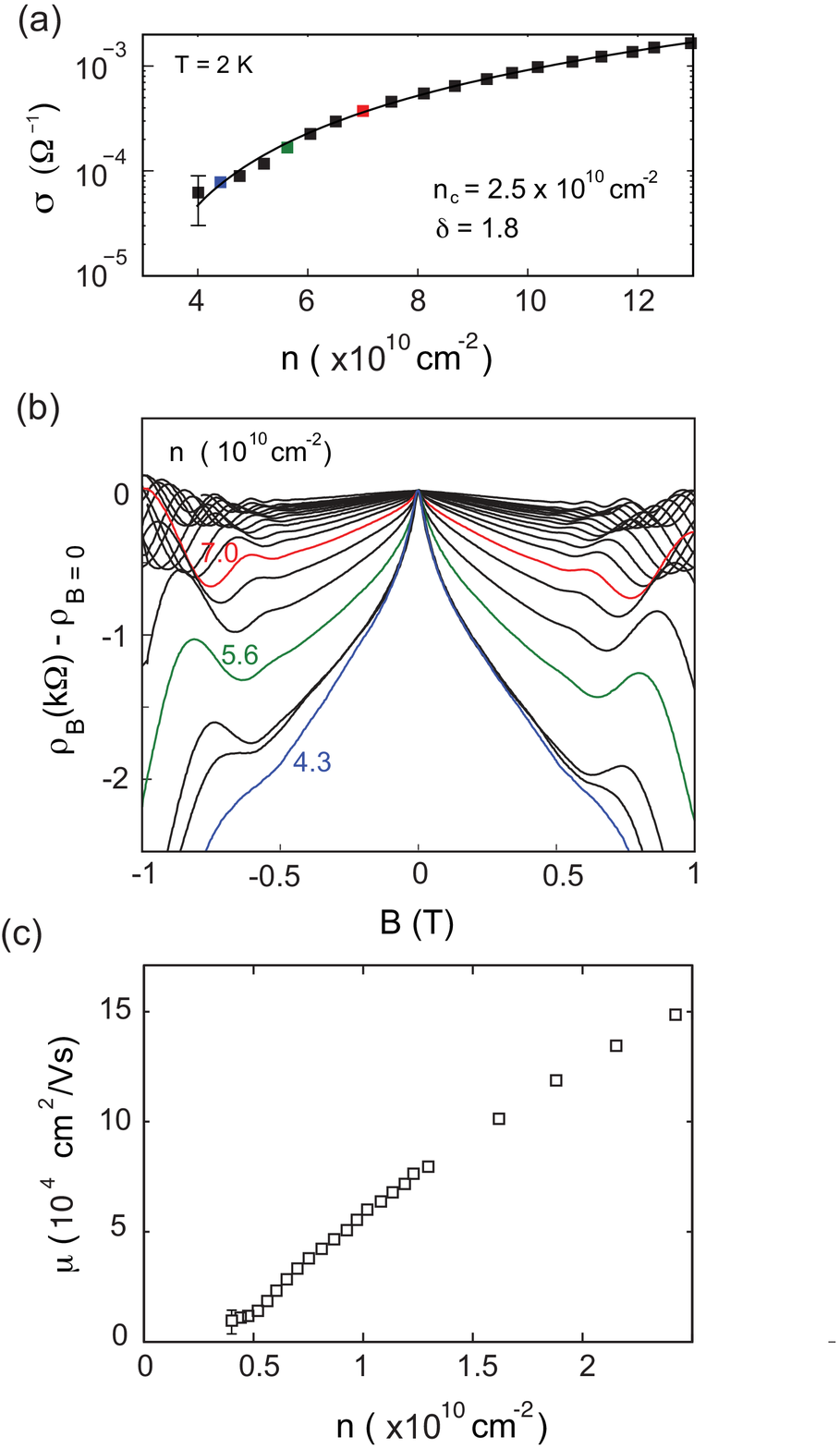}
\caption{(Color online) (a) Conductivity measured as a function of electron density (squares) and fit to Eq.~1 (solid curve). (b) Magneto-resistance in low magnetic field regime is shown. For clarity the zero magnetic field values are subtracted from each trace. (c) mobility is plotted as a function of density. The slope changes near the metal insulator transition.}
\end{figure}

Figure 3(a) shows the (weakly metallic) temperature dependence of the resistivity for $n = 3 \times 10^{11}$ cm$^{-2}$. At this density $\rho$ decreases with decreasing temperature indicating a metallic behavior. At lower densities, $n < 1 \times 10^{11}$ cm$^{-2}$, $\rho$ increases with decreasing temperature as shown in Fig. 3(b). At $n = 3.5 \times 10^{10}$ cm$^{-2}$, $\rho$ increases monotonically with decreasing temperature down to 0.5 K demonstrating an insulating behavior. The critical density which separates the insulating and metallic temperature dependences in Fig.~3 is in the vicinity of $n_{c} \sim 5 \times 10^{10}$ cm$^{-2}$. Assuming a spin degeneracy of 2, the critical resistivity (at $n=n_{c}$) corresponds to $k_{F}l \sim 0.5$ (where l is the mean free path) for our measured resistivity of $\sim 25 k\Omega$, indicating that the Ioffe-Regel criterion is approximately obeyed at $n_{c}$. Using an effective electron mass of $m^{*}_{e} = 0.03$ $m_{e}$ which is determined using the temperature dependence of Shubnikov-de Haas oscillations, we obtain a value of $r_{s}$ = 2.1 at this density. which is one of the lowest values of $r_{s}$ at which a 2D MIT has ever been reported in the literature implying that interaction effects are unlikely to be playing an important role in the current situation.

We analyze the transition from metallic to an insulating behavior using the percolation model \cite{MeirPRL99, IlaniPRL00,DasSarmaPRL05, DasSarmaHwangLiPRB13}. Based on this model, when the density of the 2DES is lowered, the system can no longer screen the random background disorder potential and disorder drives the system into an inhomogeneous potential and density landscape (i.e. puddles) with an MIT occurring when the conducting path through the puddles vanishes at the percolation threshold.  This percolation model for 2D MIT has been successfully employed before for 2D MIT in several different systems including n- and p-GaAs \cite{MeirPRL99,DasSarmaPRL05, ManfraPRL07} as well as n-Si MOSFETs \cite{TracyPRB09}. In this model the conductivity near the critical density is described by:

\begin{equation}
\sigma \sim (n-n_{c})^{\delta}
\end{equation}

where $n_{c}$ is the critical density, and $\delta$ is the critical exponent describing the transition. Using Eq.~1 we have fitted our data where the density dependence of the conductivity exhibits an exponent of $\delta = $1.79 ($\delta \sim 1+\alpha$ where $\alpha$ is the corresponding mobility exponent), which indicates that the localization is caused simply by carrier trapping at defects ~\cite{DasSarmaPRB13}. The critical density derived from the percolation is $ n_{c} \sim 2.5 \times 10^{10}$ cm$^{-2}$ which is consistent with the temperature dependence of the resistivity (Fig.~4(a)). In Fig.~4(b) we show the magneto-resistance data near zero magnetic field. For $n < 1 \times 10^{11}$ cm$^{-2}$, we observe a sharp enhancement of magneto-resistance around B = 0 T which signals possibly strong electron-electron interaction and localization effects. However the conductivity does not show a logarithmic dependence with temperature. The electron mobility also changes slope near MIT as shown in Fig.~4(c).

\begin{figure}[htp]
\centering
\includegraphics[scale=0.5]{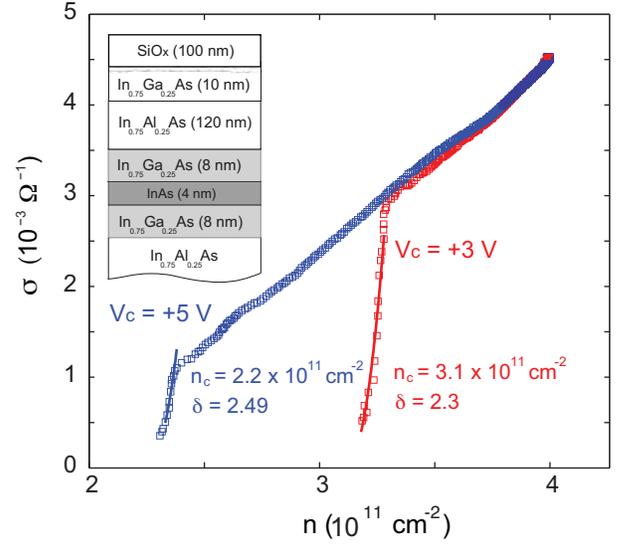}
\caption{Conductivity measured as a function of density for a structure that supports a conductive channel on the surface of the InGaAs layer (shown as inset). Sample is cooled down from room temperature with a positive bias, $V_{c}$ = +3 V (red) and +5 V (blue). The fits (solid curves) are shown for both cases near the insulating phase. }
\end{figure}

We investigated the percolation transition model further by measuring a sample with an identical structure adding a 10 nm $In_{0.75}Ga_{0.25}As$. The schematic of the growth structure is shown in the inset of Fig.~5. We find a parasitic conductive layer forming at the surface which shields the applied electric field from affecting the 2DES. However, we can reduce this screening by applying a positive bias during cool down from 300 K down to 2 K \cite{BuksPRB94}. Figure 5 summarizes our findings on two cool downs under two positive biases: $V_{c} = $ +3 V (red) and +5 V (blue). At high densities ($n>3.5 \times 10^{11}$cm$^{-2}$), we find the same exponent that fits mobility vs density, $\mu = n^{\alpha}$ with an $\alpha$ = 1.35 for both cool-downs. This large value of $\alpha$ (with the corresponding 2D conductivity exponent $\delta \sim \alpha$ + 1 = 1.35 +1 =2.35) is close to the limiting unscreened Coulomb disorder value of 1.5 \cite{DasSarmaSSC05,DasSarmaPRB13, DasSarmaPRB04} indicating the dominance of remote scattering in restricting the 2D mobility. The dominant role of remote scattering is consistent with the effective value of the dimensionless parameter in the system which is given by $k_{F}d$ where {\it d} is the average distance of the charged impurities from the 2DES. Here we have free carriers in InGaAs layer due to the Fermi level pinning at the surface \cite{TsuiPRL70, ShabaniArxivGating}. In our structure, $d \sim$ 130 nm gives the dimensionless $k_{F}d$ parameter to be around 19 for $n= 4\times10^{11}$ cm$^{-2}$ 2D carrier density. A large value of the $k_{F}d$ parameter, according to the recent theory of Das Sarma and Hwang \cite{DasSarmaPRB13}, should give an $\alpha \sim 1.5$ which is consistent with the experimental observation of $\alpha \sim 1.35$. The experimental $\alpha$ is somewhat less than the pure remote scattering theoretical $\alpha$ of 1.5 probably because the background unintentional scattering is not completely insignificant which would suppress the value of $\alpha$ somewhat since background impurity scattering typically gives $\alpha \sim 0.5$ \cite{DasSarmaPRB13, DasSarmaPRB04}. We find a critical density of $n_{c} = 2.2$ (for $V_{c}$ = +5 V) and $n_{c} = 3.1$ (for $V_{c}$ = +3 V) in units of $\times 10^{11}$ cm$^{-2}$. Although the critical density deduced from the fit to Eq.~1 near MIT varies by $\sim 1 \times 10^{11}$ cm$^{-2}$, the exponents $\alpha$ and $\delta$ show similar values for both cool-downs in Fig.~5. Positive bias changes the density of remote impurities affecting the potential landscape in the 2DES which causes a change in the critical density. Indeed for $V_{c}$ = +5 V where there are less active impurities, the critical density is lower. We note that $r_{s}$ = 1 for $V_{c}$ = +5 V and $r_{s}$ = 0.85 for $V_{c}$ = +3 V. This suggests that 2D MIT in InAs is well described by a density-inhomogeneity driven percolation transition rather than being an interaction-driven quantum phase transition. This is also consistent with our finding of the expected weak antilocalization effect in this system at higher carrier densities \cite{ShabaniUnpublished}.

In conclusion we have observed an apparent metal insulator transition in InAs heterostructures and interpreted it in the context of a percolation transition driven by the failure of screening in the low density regim. At high densities, we identifiy background charged impurities as the main scattering mechanisms limiting the 2DES mobility. We also studied the metal insulator transition by introducing remote impurities in the system establishing that the transport properties can be well-explained as arising from background or remote scattering depending on the dominance of one or the other, thus further reinforcing the percolation model for the 2D MIT behavior in the system.

Our work was supported by the Microsoft Research. 

\begin{center}
{\bf References}
\end{center}
\end{document}